\documentclass[twocolumn,showpacs,preprintnumbers,amssymb,prc]{revtex4}
\usepackage{graphicx}
\usepackage{slashbox}
\usepackage{dcolumn}
\usepackage{bm}
\usepackage{latexsym,epsfig}

\begin{document}
\title{Appraising nuclear octupole moment contributions to the hyperfine structures in $^{211}$Fr} 

\author{B. K. Sahoo \footnote{Email: bijaya@prl.res.in}}

\affiliation{Theoretical Physics Division, Physical Research Laboratory, Ahmedabad-380009, India}

\date{Received date; Accepted date}

\begin{abstract}
Hyperfine structures of $^{211}$Fr due to the interactions of magnetic dipole ($\mu$), electric quadrupole 
($Q$) and magnetic octupole ($\Omega$) moments with the electrons are investigated using the relativistic  
coupled-cluster (RCC) theory with an approximation of singles, doubles and important valence triples excitations 
in the perturbative approach. Validity of our calculations are substantiated by comparing the results with their available 
experimental values. Its $Q$ value has also been elevated by combining the measured hyperfine structure constant of 
the $7p \ ^2P_{3/2}$ state with our improved calculation. Considering the preliminary value of $\Omega$ from the 
nuclear shell-model, its contributions to the hyperfine structures up to the $7d \ ^2D_{5/2}$ low-lying states in 
$^{211}$Fr are estimated. Energy splittings of the hyperfine transitions in many states have been assessed to find out 
suitability to carry out their precise measurements so that $\Omega$ of $^{211}$Fr can be inferred from them unambiguously.  
\end{abstract}

\maketitle

\section{Introduction}
When the non-central electromagnetic fields of nuclear magnetic moments interact with the atomic electrons, they give rise 
hyperfine structures to the atomic energy levels \cite{casimir}. These energy levels are conventionally expressed in terms 
of the hyperfine structure constants, which are traditionally denoted by $A$, $B$, $C$ etc., that depend on the electron wave 
functions and nuclear moments. These constants are typically determined experimentally by measuring hyperfine structure energy 
levels for different angular momentum states of the atomic systems and fitting in a set of equations involving the corresponding 
angular momentum coefficients \cite{charles}. Among all the hyperfine structure interactions, the magnetic dipole hyperfine 
component typically contributes predominantly for a finite nuclear spin ($I$) atomic systems compared to others followed 
by the electric quadrupole hyperfine component and so on when $I> 1/2$. Owing to comparable orders of magnitudes of the higher 
multipole contributions with the systematics in the measurements of hyperfine splittings, it is extremely strenuous to estimate 
contributions due to these higher nuclear moments, especially from the nuclear octupole moment on wards, to the hyperfine 
structure. Thus, hyperfine structure constants mostly up to the electric quadrupole interactions are usually reported by the 
experimental studies. However with the advent of modern technologies, it is now possible to measure hyperfine splittings very 
precisely from which contributions until the magnetic octupole interaction are segregated in a number of systems like $^{85}$Rb 
\cite{ye}, $^{133}$Cs \cite{verginov1}, $^{137}$Ba$^+$ \cite{barrett} and $^{173}$Yb \cite{natarajan}.

On the other hand direct measurements of nuclear moments are obscured, since isolating a bare nucleus from the external 
electromagnetic fields is not a straightforward task. They are obtained either from the nuclear magnetic resonance (NMR), nuclear
quadrupole resonance (NQR) etc. measurements or by combining experimental results of $A$, $B$, $C$ etc. with their corresponding 
atomic and molecular calculations. Nuclear magnetic moment ($\mu$) obtained from an NMR measurement is generally more precise 
than its value obtained from the hyperfine structure constant due to achievement of high precision results using the NMR 
spectroscopy technique \cite{stones}. But, many of the available nuclear quadrupole moments ($Q$s) for different isotopes are 
obtained from the measured hyperfine structure constants \cite{stones}. Similarly, all the reported nuclear octupole moments 
($\Omega$s) for the above investigated atoms are inferred from the high precision measurements on hyperfine splittings 
\cite{verginov1,barrett,natarajan,verginov2}. Nevertheless there are no NMR spectroscopy of the Fr isotopes available till date,
$\mu$ and $Q$ values of $^{211}$Fr were extracted out by measuring energies of different hyperfine transitions in its ground state 
using the atomic-beam magnetic resonance (ABMR) method \cite{ekstroem} and by combining measured $B$ value of the $7p \ ^2P_{3/2}$ 
state with an atomic calculation using a lower-order relativistic many-body calculation \cite{martensson}, respectively. 

Theoretical studies of hyperfine structure constants are of immense interest because, comparison with their corresponding 
experimental results are served as bench marks to test the capability of an employed many-body method for their precise 
determination. Since accuracies in the evaluation of these quantities are very sensitive in the calculations of the atomic wave 
functions in the nuclear region, their precise estimates are the indications of the potentials of the methods to produce the 
atomic wave functions appropriately in this region. Thats why their theoretical studies have been drawn a lot of attentions in 
the context of investigating atomic parity violating (APV) effects and electric dipole moments (EDMs) due to parity and 
time-reversal symmetry violations in the atomic systems \cite{bijaya1,bijaya2}. Fr is under consideration for both the APV and 
EDM measurements \cite{sakemi,stancari,dreischuh,gomez}. In our previous work \cite{bijaya}, we had applied relativistic many-body 
methods at various approximations to determine the $A$ and $B$ constants of $^{210}$Fr and $^{212}$Fr. In that work, we had found 
that the relativistic coupled-cluster (RCC) theory at the truncation level of singles, doubles and important valence triple
excitations in the perturbative approach (CCSD$_{\text{t3}}$ method) gives rise the results within the reasonable accuracies. Since 
the aim of the present work is to estimate $\mu$ and $Q$ values of $^{211}$Fr precisely and demonstrate possible way of extracting 
$\Omega$ value of $^{211}$Fr, we carry out the calculations using the same CCSD$_{\text{t3}}$ method for this isotope here.

\section{Theory}

The Hamiltonian describing the non-central hyperfine interactions between the electrons and nucleus in an atomic system in terms of 
the tensor operators is given by \cite{charles}
\begin{eqnarray}
H_{hf} = \sum_k {\bf M}^{(k)} . {\bf T}^{(k)},
\end{eqnarray}
where ${\bf M}^{(k)}$ and ${\bf T}^{(k)}$ are the spherical tensor operators with rank ``$k$ ($>0$)" in the nuclear and electronic 
coordinates, respectively. Since these interactions are very weak, they are typically considered up to the first or at most till the 
second order perturbation. Due to coupling between the electronic ($J$) and nuclear ($I$) angular momenta, the total angular momentum 
of the hyperfine states are given by the vector sum ${\bf F}= {\bf J} \oplus {\bf I}$. Thus the hyperfine states are described by 
$F=|{\bf F}|$ and by its corresponding $M_F$ azimuthal quantum numbers; that are generally denoted by $| \gamma I J; F M_F \rangle$ 
with $\gamma$ representing the rest of the unspecified quantum numbers (most importantly the parity).

The first and the second orders changes in the energies of an atomic energy level due to the hyperfine interactions can be given 
by \cite{charles}
\begin{widetext}
\begin{eqnarray}
W_{F,J}^{(1)} &=& \sum_{k_1} (-1)^{I+J+F} \left  \{
                               \matrix
                                  {                                J & I & F \cr
                                I & J & k_1 \cr
                                 }
           \right \} \langle I||M^{(k_1)}||I \rangle \langle J||T^{(k_1)}||J \rangle 
\label{aeq}
           \end{eqnarray}
and 
\begin{eqnarray}
W_{F,J}^{(2)} &=& \sum_{J'} \frac{1}{E_J -E_{J'}} \sum_{k_1,k_2} \left  \{
                               \matrix
                                  { F & J & I \cr
                                k_1 & I & J' \cr
                                 }  \right \} \left  \{
                               \matrix
                                  { F & J & I \cr
                                k_2 & I & J' \cr
                                 }  \right \}
 \langle I||M^{(k_1)}||I \rangle \langle I||M^{(k_2)}||I \rangle \langle J'||T^{(k_1)}||J \rangle \langle J'||T^{(k_2)}||J \rangle ,
 \label{beq}
\end{eqnarray}
\end{widetext}
where $M^{(k)}$ and $T^{(k)}$ are the magnitudes of the ${\bf M}^{(k)}$ and ${\bf T}^{(k)}$ tensors, respectively, the allowed
values of ranks $k_1$ and $k_2$ in the above multipole expansions depend on the selection rules of Sixj angular momentum factors 
and $E_J$ is the energy of an atomic state with angular momentum $J$. In the present interest of study, we restrict the multipole 
values only to $k_{1,2}=1,2,3$. Definitions of the nuclear momentums are given by
\begin{eqnarray}
\langle II|M^{(1)}|II\rangle &=& \mu = \mu_N \mu_I , \nonumber \\
\langle II|M^{(2)}|II\rangle &=& \frac {1}{2} Q, \nonumber \\
\langle II|M^{(3)}|II\rangle &=& - \Omega = -\mu_N \Omega_I \ \text{etc.,}
\end{eqnarray}
where $\mu_I$ and $\Omega_I$ are the dimensionless nuclear dipole and octupole moments respectively, and $\mu_N=\frac{1}{2M_p c}$
is the nuclear magneton with the proton mass $M_p$. Nuclear shell-model yields expressions for $\mu_I$, $Q$ and $\Omega_I$ 
as \cite{charles}
\begin{eqnarray}
 \mu_I &=& - \langle I I | r^1 C_0^{(1)}(\theta, \phi) (\vec \nabla \times \vec {\bf D}) | II \rangle/\mu_N \nonumber \\ 
  &=& \left\{\begin{array}{ll}
      \displaystyle
      g_l +(g_s-g_l)/2I & \mbox{for } I= l+1/2
         \\ [2ex]
      \displaystyle
       g_l -(g_s-g_l)/(2I+2) & \mbox{for } I= l-1/2, 
    \end{array}\right.   \\
Q &=& \langle I I | g_l r^2 C_0^{(2)}(\theta, \phi) | II \rangle \nonumber \\
 &=& - \frac{2I-1}{2I+2} g_l \langle r^2 \rangle 
\end{eqnarray}
and
\begin{eqnarray}
 \Omega_I &=& \langle I I | r^3 C_0^{(3)}(\theta, \phi) (\vec \nabla \times \vec {\bf D}) | II \rangle/\mu_N \nonumber \\ 
&=& \frac{3(2I-1)}{2(2I+4)(2I+2)} \langle r^2 \rangle \nonumber \\
  && \times \left\{\begin{array}{ll}
      \displaystyle
      (I+2)[(I-3/2)g_l +g_s] & \mbox{for } I= l+1/2
         \\ [2ex]
      \displaystyle
        (I-1)[(I+5/2)g_l -g_s] & \mbox{for } I= l-1/2, 
    \end{array}\right.
\end{eqnarray}
with the nuclear magnetization density $\vec {\bf D}$,  $\langle r^2 \rangle$ is the mean square radius of the nucleus 
and $g_l$ and $g_s$ are the orbital and spin gyromagnetic constants for the odd nucleon of the nucleus (as applicable in  
case of $^{211}$Fr). Values of the nuclear spin $I (= l-1/2)=9/2$ \cite{ekstroem1} and moment $\mu_I = 4.00(8)$ \cite{ekstroem} 
of $^{211}$Fr indicates that this isotope has an odd-proton in the $\pi h_{9/2}$ level following the nuclear shell-model. Hence,
the nuclear moments of $^{211}$Fr can be estimated using the shell-model by substituting values as $g_l \approx 1.16$ and 
$g_s \approx 5.5857$ \cite{NIST1} and $ \langle r^2 \rangle  \approx 0.308 \ b$ \cite{angeli}.

Rewriting Eq. (\ref{aeq}) in terms of contributions from the individual multipoles, it yields 
\begin{eqnarray}
W_{F,J}^{(1)} &=& W_{F,J}^{M1} + W_{F,J}^{E2} + W_{F,J}^{M3},
\end{eqnarray}
where $W_{F,J}^{M1}$, $W_{F,J}^{E2}$, and $W_{F,J}^{M3}$ are the contributions due to the magnetic dipole (M1, $k=1$), 
electric quadrupole (E2, $k=2$) and magnetic octupole (M3, $k=3$) interactions, respectively, which are given by
\begin{eqnarray}
W_{F,J}^{M1} &=& A {\bf I.J} \\
W_{F,J}^{E2} &=& B \frac {3({\bf I.J})^2 + \frac {3}{2}({\bf I.J})- I(I+1)J(J+1)}{2I(2I-1)J(2J-1)} 
\end{eqnarray}
and
\begin{eqnarray}
W_{F,J}^{M3} &=& C \frac{1}{[I(I-1)(2I-1)J(J-1)(2J-1)]} \nonumber \\ && \times \big \{10({\bf I.J})^3+20({\bf I.J})^2  +2({\bf I.J}) 
\nonumber \\ && \times \big [-3I(I+1)J(J+1)+ I(I +1) \nonumber \\ && +J(J+1)+3 \big ] -5I(I+1)J(J+1)\big \}.
\end{eqnarray}
Here the constants are defined as
\begin{eqnarray}
A&=& g_I \frac {\langle J||T^{(1)}||J\rangle}{\sqrt{J(J+1)(2J+1)}} \label{eqna} \\
B&=& 2eQ [ \frac {J(2J-1)}{(J+1)(2J+1)(2J+3)}]^{1/2} \langle J||T^{(2)}||J\rangle \label{eqnb} \ \ \ \ \ \\
\text{and} && \nonumber \\
C&=& - \ \Omega_I \ \frac {\langle J||T^{(3)}||J\rangle}{\sqrt{J(J+1)(2J+1)}} \label{eqnc} 
\end{eqnarray}
with $g_I=\frac{\mu_I}{I}$.

Since the primary objective of this work is to estimate $C$ values of atomic states and to know typical order of magnitudes 
of their contributions to the hyperfine splittings in $^{211}$Fr so that measurements on these splittings can be carried out 
within the precision from which we can infer $\Omega$ of $^{211}$Fr reliably. In this scenario, it is also imperative to know order of 
magnitudes due to the dipole-dipole and dipole-quadrupole second order hyperfine interactions as their contributions are generally
in the same order with the magnetic octupole contribution. Expressions for the second order energy shifts are given by
\begin{eqnarray}
W_{F,J}^{(2)} &=& W_{F,J}^{M1-M1}  + W_{F,J}^{M1-E2} \nonumber \\ &=& \left | \left  \{
                               \matrix
                                  { F & J & I \cr
                                1 & I & J' \cr
                                 }  \right \}  \right |^2 \eta  \nonumber \\
   && + \left  \{
                    \matrix
                     { F & J & I \cr
                      1 & I & J' \cr
                     }  \right \}  \left  \{
                    \matrix
                     { F & J & I \cr
                      2 & I & J' \cr
                     }  \right \} \zeta,
\end{eqnarray}
with $\eta=\frac{(I+1)(2I+1)}{I} \mu_I^2 \frac{|\langle J'||T^{(k1)}||J \rangle|^2}{E_J -E_{J'}}$ and 
$\zeta=\frac{(I+1)(2I+1)}{I} \sqrt{\frac{2I+3}{2I-1}} \mu_I Q \frac{\langle J'||T^{(k1)}||J \rangle \langle J'||T^{(k2)}||J \rangle}{E_J -E_{J'}}$.

The expressions for the tensor operators for the electronic part in the above expressions given by \cite{charles}
\begin{eqnarray}
T_q^{(1)} &=& \sum_j t_q^{(1)}(r_j) = \sum_j -ie \mu_N \sqrt{8\pi/3} r_j^{-2} {\bf \alpha_j.Y_{1q}^{(0)}(r_j) } \ \ \\
T_q^{(2)} &=& \sum_j t_q^{(2)}(r_j) = \sum_j -ie  r_j^{-3} C_q^{(2)}(r_j) \ \ \text{and} \\
T_q^{(3)} &=& \sum_j t_q^{(3)}(r_j) = \sum_j -ie \mu_N \sqrt{8\pi/3} r_j^{-4} {\bf \alpha_j.Y_{3q}^{(0)}(r_j) } , \ \ \ 
\end{eqnarray}
whose single particle matrix elements are
\begin{eqnarray}
\langle \kappa m | t_q^{(1)} |\kappa'm' \rangle &=& - \mu_N \langle -\kappa m | C_q^{(1)} |\kappa'm' \rangle(\kappa+\kappa')
 \nonumber \\ && \int_0^{\infty} dr \frac {1}{r^2} (P_{\kappa} Q_{\kappa'} + P_{\kappa} Q_{\kappa'}) \label{eqa} \\
\langle \kappa m | t_q^{(2)} |\kappa'm' \rangle &=& - \langle \kappa m | C_q^{(2)} |\kappa'm' \rangle
 \nonumber \\ && \int_0^{\infty} dr \frac{1}{r^3} (P_{\kappa} P_{\kappa'} + Q_{\kappa} Q_{\kappa'})  \label{eqb}
 \end{eqnarray}
 and
 \begin{eqnarray}
\langle \kappa m | t_q^{(3)} |\kappa'm' \rangle &=& - \frac {\mu_N}{3} \langle -\kappa m | C_q^{(3)} |\kappa'm' \rangle(\kappa+\kappa')
\nonumber \\ && \int_0^{\infty} dr \frac {1}{r^4} (P_{\kappa} Q_{\kappa'} + P_{\kappa} Q_{\kappa'}) \label{eqc}.
\end{eqnarray}
In these expressions $\kappa_i$ and $j_i$ are the angular momentum quantum numbers of $i^{th}$
Dirac orbital with large $P_i$ and small $Q_i$ radial components and the matrix element of the
Racah operator is given by
\begin{eqnarray}
\langle \kappa m | C_q^{(k)}|\kappa'm' \rangle = (-1)^{j-m} \left ( \matrix {
                                       j & k & j' \cr
                                       -m & q & m' \cr
                                          }
                            \right ) \langle \kappa || C^{(k)}||\kappa' \rangle \ \ \ \ \
\end{eqnarray}
with its reduced matrix element
\begin{eqnarray}
\langle \kappa || C^{(k)}||\kappa' \rangle &=& (-1)^{j+1/2} \sqrt{(2j+1)(2j'+1)} \nonumber \\ &&
                          \left ( \matrix {
                              j & k & j' \cr
                              1/2 & 0 & -1/2 \cr
                                       }
                            \right ) \pi(l,k,l') \ \ \
\end{eqnarray}
that satisfies the condition
\begin{eqnarray}
  \pi(l,k,l') &=&
  \left\{\begin{array}{ll}
      \displaystyle
      1 & \mbox{for } l+k+l'= \mbox{even}
         \\ [2ex]
      \displaystyle
        0 & \mbox{otherwise.}
    \end{array}\right.
\end{eqnarray}

\section{A brief description of the CCSD$_{\text{t3}}$ method}

The wave function ($|\Psi_n \rangle$) of an atomic state corresponding to the closed-shell configuration $[6p^6]$ and a valence 
orbital $n$ in Fr in the RCC theory framework is expressed as \cite{bijaya,sahoo}
\begin{eqnarray}
 |\Psi_n \rangle = e^T \{1+S_n \} |\Phi_n \rangle,
 \label{eqcc}
\end{eqnarray}
where $T$ and $S_n$ are the excitation operators involving core and core-valence electrons, respectively, with the reference 
state $|\Phi_0 \rangle$, which is obtained using the Dirac-Fock (DF) method in the present work. In the CCSD$_{\text{t3}}$ method, 
the RCC excitation operators are given by \cite{bijaya}
\begin{eqnarray}
T = T_1 + T_2  \ \ \ \ \text{and} \ \ \ \ S_n = S_{1n} + S_{2n} ,
\end{eqnarray}
where the subscripts 1 and 2 represent for the single and double excitations.

The hyperfine structure constants are determined by evaluating the expression
\begin{eqnarray}
\frac{\langle \Psi_n | O | \Psi_n \rangle} {\langle \Psi_n| \Psi_n \rangle}  
&=& \frac{\langle \Phi_n | \{1+ S_n^{\dagger}\} 
e^{T^{\dagger}} O e^T \{1+S_n\} | \Phi_n \rangle} {\langle \Phi_n| \{1+ S_n^{\dagger}\} 
e^{T^{\dagger}} e^T \{1+S_n\} | \Phi_n \rangle }, \nonumber \\
\label{prpeq}
\end{eqnarray}
in the approach as has been described in \cite{bijaya}, where $O$ represents one of the hyperfine interaction operators. 
Contributions from the important triple excitations are considered perturbatively through the above expression. 

\section{Discussions and Results}

Before perusing in estimating $W_{F,J}^{M3}$ contributions for different states, it would be appropriate to test the accuracies
in our calculations of the hyperfine structure constants with respect to their experimental values. In order to estimate $A$ and 
$B$ values theoretically, it also requires knowledge of $\mu$ and $Q$ values. Ekstr\"om {\it et al} had measured hyperfine 
splittings between at least 14 sub-states in the ground state of $^{211}$Fr employing the ABMR technique and adopted a least-square fit 
on them to separate out the electronic and nuclear parts of the hyperfine structure from which they had extracted out its $\mu$ value as 
4.00(8) $\mu_N$ \cite{ekstroem}. This value seems to be very precise, but $Q$ of $^{211}$Fr was first estimated by Heully and M{\aa}rtensson-Pendrill
as $-0.24$b combining the measured $B$ value $-55.3(3.4)$ MHz of the $7p \ ^2P_{3/2}$ state reported in Ref. \cite{liberman} with 
their calculation of $B/Q$ as 231.0 MHz/b using a relativistic many-body method that takes into account only the DF contribution and 
core-polarization effects to all orders \cite{martensson}. Later Coc {\it et al} reported another measurement of $B$ value of this state 
as $-51.0(7.0)$ and revised the $Q$ value to $-0.19(3)$b \cite{coc}, but they had still used the same $B/Q$ value of Heully and M{\aa}rtensson-Pendrill 
and accounts uncertainty only from the experimental $B$ value. It is worth mentioning here that uncertainty in the earlier measured $B$
value by Liberman {\it et al} is comparatively smaller than that of the result reported by Coc {\it et al}. Nevertheless, our employed 
CCSD$_{\text{t3}}$ method takes care of contributions from the DF method, pair-correlation effects to all orders and core-polarization 
effects to all orders \cite{sahoo,baphyp}. Thus to improve accuracy in the $Q$ value further, we combine our CCSD$_{\text{t3}}$ value for $B/Q$, 
which is 259.73 MHz/b, with the Liberman {\it et al} measurement (which seems to be more precise) of $B$ as $-55.3(3.4)$ MHz of the 
$7p \ ^2P_{3/2}$ state and obtain the new $Q$ value as $-0.21(2)$b. The reason for which Coc {\it et al} had got lower $Q$ value owes
to smaller absolute $B$ value measured in their experiment.

\begin{table}[t]
\caption{Comparison of the nuclear magnetic moment ($\mu_I$), quadrupole moment $Q$ in unit of barn ($b$) and octupole moment 
($\Omega_I$) in unit of $b$ for $^{211}$Fr between the nuclear shell-model calculations and estimations from the atomic studies of 
hyperfine structures.}\label{tab1}
\begin{ruledtabular}
\begin{tabular}{lccc}
Moment  &  Shell-model & Atomic study & Reference \\
\hline \\
 $\mu_I$  &  3.41  &  4.00(8) & \cite{ekstroem} \\
 $Q$ &    $-0.26$     &  $-0.19(3)$ & \cite{ekstroem}  \\
     &                &  $-0.24$    & \cite{martensson} \\
     &         &  $-0.21(2)$ & This work \\
 $\Omega_I$ & 0.229  &  Unavailable & \\ 
\end{tabular}
\end{ruledtabular}
\end{table}
In Table \ref{tab1}, we now give the values of nuclear moments of $^{211}$Fr estimated using the shell-model and compare them against 
the precise values obtained from the atomic studies. This comparison demonstrates that the nuclear shell-model values for $\mu$ and 
$Q$ differ only by 15-25\% from the atomic results. Since till date hyperfine splittings in $^{211}$Fr are not yet measured very 
precisely to be able to infer $\Omega$ (or $\Omega_I$) value from those measurements, the preliminary value obtained from the 
shell-model can be used to estimate contributions from the octupole component to the hyperfine splittings by combining with our 
calculations so that some ideas about how precisely it is required to measure the hyperfine splittings to infer the $\Omega$ value 
of $^{211}$Fr unambiguously can be gauge. 

\begin{table*}
\caption{Theoretical and experimental results of $A$, $B$ and $C$ in MHz. Results under the head of DF and CCSD$_{\text{t3}}$ methods
are from the present work. References for the other results are given at the bottom of the table.} 
\begin{ruledtabular}
\begin{tabular}{l|ccc|ccc|cc}
  & \multicolumn{3}{c|}{$A$ values}  & \multicolumn{3}{c|}{$B$ values}  &  \multicolumn{2}{c}{$C \times 10^{4}$ values} \\
  \cline{2-4} \cline{5-7} \cline{8-9}\\
State  & CCSD$_{\text{t3}}$ & Others & Experiment & CCSD$_{\text{t3}}$ & Others & Experiment & DF & CCSD$_{\text{t3}}$\\
 \hline
                   &         &         &             &   &   &   &   & \\
 $7s \ ^2S_{1/2}$  & 8786.20  & 8833.0$^a$& 8713.9(8)$^b$  &   &   &   &   &    \\
                   &         &  9027.04$^c$& 8698.2(10.5)$^d$ &   &   &   &   & \\
\vspace{0.4mm} 
 $7p \ ^2P_{1/2}$  & 1137.40 & 1162.1$^a$ & 1139.2(14)$^b$  &  &   &   &   & \\
                   &         & 1125.12$^c$& 1142.1(2)$^e$ &  &   &   &   & \\
                   &         & 778.0$^f$&   &  &   &   &   & \\
\vspace{0.4mm} 
 $7p \ ^2P_{3/2}$  & 92.68  & 91.80$^a$ &  94.9(3)$^b$ &  $-54.54$ & $-48.51^f$ & $-51.0(7.0)^b$  & $5.99$ & $11.0$ \\
                   &         & 102.34$^c$&  94.7(2)$^d$ &        &       &  $-55.3(3.4)^d$       &          & \\
                   &         & 85.5$^f$&   &        &       &         &          & \\
 \vspace{0.4mm} 
 $6d \ ^2D_{3/2}$  & 75.72   &          &     & $-20.71$ & &  & $0.62$ & $1.76$ \\
\vspace{0.4mm} 
 $6d \ ^2D_{5/2}$  & $-52.28$&    &          & $-27.59$ & & & $0.18$ &  $-3.26$ \\
\vspace{0.4mm} 
 $8s \ ^2S_{1/2}$  & 1912.64 & 1923.3$^a$ &   &  &   &   &   & \\
                   &         & 1971.96$^c$ &   &  &   &   &   & \\
\vspace{0.4mm} 
 $8p \ ^2P_{1/2}$  & 357.36  & 362.91$^a$&  358.0(4)$^g$ &  &   &   &   & \\
                   &         & 364.14$^c$&   &  &   &   &   & \\
\vspace{0.4mm} 
 $8p \ ^2P_{3/2}$  & 31.12   &  30.41$^a$& 31.47(4)$^g$  & $-17.64$ &  & & $2.20$ & $3.68$ \\
                   &         &  35.24$^c$&   &  &   &   &   & \\
 \vspace{0.4mm} 
 $7d \ ^2D_{3/2}$  & 27.48   &     &        & $-6.41$ &  & & $0.31$ & $0.55$ \\
\vspace{0.4mm} 
 $7d \ ^2D_{5/2}$  & $-14.48$ &    &         & $-8.38$ & & & $0.09$ & $-0.73$ \\
\end{tabular}
\end{ruledtabular}
\label{tab2}
\begin{tabular}{ll}
References: & $^a$\cite{safronova}.\\ 
            & $^b$\cite{coc}. \\
            & $^c$\cite{dzuba}.\\
            & $^d$\cite{liberman}.\\
            & $^e$\cite{grossman}.\\
            & $^f$\cite{martensson} (new $Q$ value is multiplied to get $B$). \\
            & $^g$\cite{touchard}.\\
\end{tabular}
\end{table*}

Using the CCSD$_{\text{t3}}$ calculations for $A/\mu$ and $B/Q$ and considering the precise values of $\mu$ and $Q$ values of 
$^{211}$Fr from Table \ref{tab1}, we evaluate theoretical values of $A$ and $B$. These values are given in 
Table \ref{tab2} along with the results from the previous calculations and available experimental results. Comparison
of our calculations with the experimental results give some indication on the accuracies of our results. Calculations
for some of the states are reported by Safronova {\it et al} using a similar method like ours but keeping only the linear
terms in the RCC theory (SDpT method) \cite{safronova}. Dzuba {\it et al} had also employed a restricted Hartree-Fock 
method in the relativistic framework and incorporated correlation effects using many-body perturbation theory to investigate 
correlation effects in the hyperfine structure constants of a few low-lying states of $^{211}$Fr \cite{dzuba}. Heully and 
M{\aa}rtensson-Pendrill had used a relativistic many-body perturbation method treating polarization effects to all orders 
\cite{martensson}. Our calculations match well with the calculations of Safronova {\it et al} and Dzuba {\it et al}, but 
calculations by Heully and M{\aa}rtensson-Pendrill differ significantly from all the other calculations. The reason may be 
owing to predominant contributions from the pair-correlation effects to the hyperfine structure, which are missed out in the
method employed by Heully and M{\aa}rtensson-Pendrill. In fact, theoretical results of $A$ from the CCSD$_{\text{t3}}$ method 
seem to be more accurate among all the calculations. This suggests that our $B/Q$ and $C/\Omega$ calculations are also reliable 
enough to be used for inferring $Q$ and $\Omega$ values combined with their corresponding $B$ and $C$ values (may be measured
in future). So this justifies why our extracted new $Q$ value is more valid than the previously estimated values. Now combining 
the new $Q$ value with the CCSD$_{\text{t3}}$ results and calculation by Heully and M{\aa}rtensson-Pendrill of $B/Q$, we get the 
$B$ values for all the considered states of $^{211}$Fr. In fact, we have calculated these quantities for many states as gives
freedom to the experimentalists to select suitable states as per their choices to measure the hyperfine splittings within the 
required precision so that $\Omega$ of $^{211}$Fr can be inferred. Though experimental results of $A$ for many states are not 
available to verify their accuracies, but good agreements for the states for which measurements are available provide
some confidence on their reliabilities. Moreover, consistencies among the calculated results results also imply that our 
calculations are quite accurate. On this basis, we also expect that our calculations for $B/Q$ and $C/\Omega$ are reasonably
accurate. 

\begin{table}[t]
\caption{Important off-diagonal matrix elements in MHz among the fine structure partners obtained using our DF and CCSD$_{\text{t3}}$
methods. To determine the conjugate matrix elements between the $J$ and $J'$ states from these values, one needs to multiply by the 
phase factors $(-1)^{J-J'}$.}
\begin{ruledtabular}
\begin{tabular}{lcc}
 Off-diagonal Matrix & DF & CCSD$_{\text{t3}}$ \\
\hline \\
$\langle 7p_{1/2} || T^{(1)} ||7p_{3/2} \rangle$ & $-67.42$ & $-32.78$  \\
$\langle 8p_{1/2} || T^{(1)} ||8p_{3/2} \rangle$ & $-24.35$ & $-10.44$ \\
$\langle 6d_{3/2} || T^{(1)} ||6d_{5/2} \rangle$ & $-16.12$ & $-790.87$ \\
$\langle 7d_{3/2} || T^{(1)} ||7d_{5/2} \rangle$ & 7.93 & 238.11  \\
$\langle 7p_{1/2} || T^{(2)} ||7p_{3/2} \rangle$ & $-750.86$ & $-1588.40$   \\
$\langle 8p_{1/2} || T^{(2)} ||8p_{3/2} \rangle$ & $-271.23$ & $-501.30$ \\
$\langle 6d_{3/2} || T^{(2)} ||6d_{5/2} \rangle$ & $-63.01$ & $-253.07$  \\
$\langle 7d_{3/2} || T^{(2)} ||7d_{5/2} \rangle$ & 31.03 &  76.17  \\
\end{tabular}
\end{ruledtabular}
\label{tab3}
\end{table}

 Roles of the correlation trends in the evaluation of the $A/\mu$ and $B/Q$ values both in $^{210}$Fr and $^{212}$Fr isotopes 
were already demonstrated explicitly by us recently in \cite{bijaya}; we also observe the similar trends in the $^{211}$Fr though 
there are slight changes in the results owing to different nuclear structure. Just to gain some insights into the roles of the 
correlation effects in the evaluations of the $C/\Omega$ values, we have given the estimated $C$ values from both the DF and 
CCSD$_{\text{t3}}$ methods in Table \ref{tab2} using $\Omega_I=0.229$b from the nuclear shell-model. Differences between these 
results show contributions from the electron correlation effects captured by the CCSD$_{\text{t3}}$ method in the evaluations 
of $C/\Omega$. We also observe that the trend of the correlation effects in the evaluations of $C$ and $A$ are almost similar. 
In both the cases, the core-polarization effects in the $D_{5/2}$ states are found to be extremely large and opposite than 
their DF results. This may be due to the fact that single particle expressions given in Eqs. (\ref{eqa}) and (\ref{eqc}) 
are similar except different powers on $r$. Now from this analysis, it is obvious that the $C$ values are about at least four 
orders smaller than the $A$ and $B$ values in $^{211}$Fr. This means that measurements of the hyperfine splittings in the $J>1/2$ 
states of this isotope need to be measured till the fifth decimal place accuracies in case we aim to infer $\Omega$ from these 
measurements. Also comparing to the magnitudes of the estimated $C$ values, we find $C$ values are in particularly large in the 
$7p \ ^2P_{3/2}$, $8p \ ^2P_{3/2}$ and $6d \ ^2D_{5/2}$ states. 

Hyperfine energy level of a state cannot be measured directly, but in practice their differences are being measured. Thus, it is 
essential to identify suitable hyperfine transitions among as many as sub-states possible to extract out contributions from the 
individual multipole precisely. Keeping this in mind, we give explicit expressions with appropriate angular momentum coefficients 
for the energy differences ($\delta W_J^{F-F'}$) between different hyperfine momenta (say $F$ and $F'$) for each $J-$ symmetry states
in appendices. Although expressions for the $J=1/2$ states may not be required in the estimations of the $C$ values, but they can be 
used to eliminate contributions from the off-diagonal matrix elements appearing through the second-order effects among the fine
structure partners. It is also obvious that these off-diagonal elements are one of the the major systematics in the
extractions of the $C$ values from the measured hyperfine splittings. Therefore, it is important to determine the off-diagonal 
matrix elements of the hyperfine interaction operators accurately. For the same purpose, we have also calculated these quantities using the DF 
and CCSD$_{\text{t3}}$ methods and give them in Table \ref{tab3}. Large differences between the results from these two methods imply
that proper inclusion of the correlation effects are also crucial for their accurate evaluations. Though our derived expressions
for the different $J$ states will be useful for estimating the $C$ values provided the hyperfine splittings can be measured very high 
precisely, however we would like to point out from the associated large angular momentum coefficients from the derivations given 
in the appendices that the energy differences $\delta W_{np3/2}^{6-3}$ and $\delta W_{nd5/2}^{7-2}$ of the $np \ ^2P_{3/2}$ and 
$nd \ ^2D_{5/2}$ hyperfine states seem to be more suitable for the unambiguous extractions of the $C$ values as they are free from 
the off-diagonal matrix element contributions.

\section{Conclusion}

We have studied hyperfine structures of many low-lying states in $^{211}$Fr using a relativistic coupled-cluster theory in the 
singles, doubles and partial triples excitations approximation. Its nuclear quadrupole momentum value has been revised by combining 
our atomic calculations with the measured $B$ value of the $7p \ ^2P_{3/2}$ state. We also find our calculated $A$ values are in
very good agreement with the available experimental results. By estimating $\Omega$ value from the nuclear shell-model and using 
our calculations of $C/\Omega$, preliminary values of $C$ up to the $7d \ ^2D_{5/2}$ low-lying state in $^{211}$Fr are evaluated.
We also give the off-diagonal matrix elements to determine the second order energy shifts due to the dipole-dipole and 
dipole-quadrupole hyperfine interactions, which are typically in the same order of magnitudes compared to the contributions 
from the $C$ values. From all these analysis, we find that hyperfine energy differences between the $F=6$ to $F=3$ hyperfine levels
in the $np \ ^2P_{3/2}$ states and between the $F=7$ to $F=2$ hyperfine levels in the $nd \ ^2D_{5/2}$ states are most
suitable to infer their corresponding $C$ values unambiguously; thusly $\Omega$ of $^{211}$Fr. Nevertheless, our 
derivations on the hyperfine splittings for all the important low-lying states will be very useful when the measurements
are carried over to extract out the $A$, $B$ and $C$ constants in the hyperfine transitions of $^{211}$Fr.

\section*{Acknowledgement}
This work was supported partly by INSA-JSPS under project no. IA/INSA-JSPS Project/2013-2016/February 28,2013/4098. 
Computations were carried out using the 3TFLOP HPC cluster at Physical Research Laboratory, Ahmedabad.

\appendix

\section*{Appendices}

Here, we express hyperfine structure splittings of different atomic states in terms of $A$, $B$, $C$ and the 
corresponding off-diagonal $\eta$ and $\zeta$ coefficients. Splittings for the states with same $J$ and parity but having different 
principal quantum numbers, denoted by an index $n$, are given in general forms. For $^{211}$Fr, hyperfine state angular momenta 
are determined by considering the nuclear spin $I=9/2$. Again, we account off-diagonal contributions only from the fine structure 
splittings and neglect contributions from other higher states owing to large energy denominators associated with those intermediate
states.

\section{$[(n-1)p^6] \ ns ^2S_{1/2}$ state}

For the $ns \ ^2S_{1/2}$ states, $J=1/2$. Thus, it yields $F=4,5$ and
\begin{eqnarray}
W_{4,ns}^{(1)} &=& -\frac{11}{4} A^{ns} \ \ \text{and} \ \  W_{5,ns}^{(1)} = \frac{9}{4} A^{ns} . \ \ \
\end{eqnarray}
Using these relations, it yields
\begin{eqnarray}
\delta W_{ns}^{5-4} = W_{5,ns}^{(1)} - W_{4,ns}^{(1)} = 5  A^{ns}.
\end{eqnarray}

\section{$[(n-1)p^6] \ np ^2P_{1/2}$ state}
In this case also, we have $J=1/2$ and $F=4,5$. Thus, the first order effects are given by
\begin{eqnarray}
W_{4,np1/2}^{(1)} &=& -\frac{11}{4} A^{np1/2} \ \ \text{and} \ \  W_{5,np1/2}^{(1)} = \frac{9}{4} A^{np1/2} . \ \ \ \
\end{eqnarray}

The $np_{1/2}$ states can have second order effect owing to small energy differences with their fine structure 
levels and can be given by 
\begin{eqnarray}
W_{4,np_{1/2}}^{(2)} &=& (11/3) g_I^2 \frac{|\langle J_{np3/2} || T^{(1)}_e || J_{np1/2} \rangle|^2} {E_{np1/2}-E_{np3/2}} \nonumber \\
  && - (33/2) \sqrt{1/15} g_I Q \langle J_{np3/2} || T^{(1)}_e || J_{np1/2} \rangle \nonumber \\ && \times 
  \frac { \langle J_{np3/2} || T^{(2)}_e || J_{np1/2} \rangle} {E_{np1/2}-E_{np3/2}}
\end{eqnarray}
and 
\begin{eqnarray}
W_{5,np{1/2}}^{(2)} &=& (9/2) g_I^2 \frac{|\langle J_{np3/2} || T^{(1)}_e || J_{np1/2} \rangle|^2} {E_{np1/2}-E_{np3/2}} \nonumber \\
  && + (9/2) \sqrt{3/5} g_I Q \langle J_{np3/2} || T^{(1)}_e || J_{np1/2} \rangle \nonumber \\ 
 && \times \frac{ \langle J_{np3/2} || T^{(2)}_e || J_{np1/2} \rangle} {E_{np1/2}-E_{np3/2}}.
\end{eqnarray}
This gives us
\begin{eqnarray}
\delta W_{np1/2}^{5-4} &=& W_{5,np1/2}^{(1)} - W_{4,np1/2}^{(1)} + W_{5,np1/2}^{(2)} - W_{4,np1/2}^{(2)} \nonumber \\
   &=& 5 A^{np1/2} + (5/6) g_I^2  \frac{|\langle J_{np3/2} || T^{(1)}_e || J_{np1/2} \rangle|^2} {E_{np1/2}-E_{np3/2}} \nonumber \\
   && +10 \sqrt{3/5}  g_I Q \langle J_{np3/2} || T^{(1)}_e || J_{np1/2} \rangle \nonumber \\ && \times
 \frac{ \langle J_{np3/2} || T^{(2)}_e || J_{np1/2} \rangle} {E_{np1/2}-E_{np3/2}}.
\end{eqnarray}

\section{$[(n-1)p^6] \ np ^2P_{3/2}$ state}
Considering $J=3/2$ and $I=9/2$, we have $F=3,4,5,6$. This gives us the first order splittings as
\begin{eqnarray}
W_{3,np3/2}^{(1)} &=& -(33/4) A^{np3/2} + (11/24) B^{np3/2} \nonumber \\ && - (143/42) C^{np3/2}, \nonumber \\ 
W_{4,np3/2}^{(1)} &=& -(17/4) A^{np3/2} - (5/24) B^{np3/2} \nonumber \\ && + (13/2) C^{np3/2}, \nonumber \\ 
W_{5,np3/2}^{(1)} &=& (3/4) A^{np3/2} - (5/12) B^{np3/2} \nonumber \\ && - (13/3) C^{np3/2}, \nonumber \\
W_{6,np3/2}^{(1)} &=& (27/4) A^{np3/2} + (1/4) B^{np3/2} \nonumber \\ && + C^{np3/2}
\end{eqnarray}
and the second order splittings are
\begin{eqnarray}
W_{3,np{3/2}}^{(2)} &=& 0, \nonumber \\
W_{4,np{3/2}}^{(2)} &=& (11/3) g_I^2 \frac{|\langle J_{np1/2} || T^{(1)}_e || J_{np3/2} \rangle|^2} {E_{np3/2}-E_{np1/2}} \nonumber \\
  && - (33/2)\sqrt{1/15} g_I Q \langle J_{np1/2} || T^{(1)}_e || J_{np3/2} \rangle \nonumber \\ && \times 
  \frac { \langle J_{np1/2} || T^{(2)}_e || J_{np3/2} \rangle} {E_{np3/2}-E_{np1/2}}, \nonumber \\  
  W_{5,np{3/2}}^{(2)} &=& (9/2) g_I^2 \frac{|\langle J_{np1/2} || T^{(1)}_e || J_{np3/2} \rangle|^2} {E_{np3/2}-E_{np1/2}} \nonumber \\
  && + (9/2) \sqrt{3/5} g_I Q \langle J_{np1/2} || T^{(1)}_e || J_{np3/2} \rangle \nonumber \\ && \times 
  \frac { \langle J_{np1/2} || T^{(2)}_e || J_{np3/2} \rangle} {E_{np3/2}-E_{np1/2}}, \nonumber \\
W_{6,np{3/2}}^{(2)} &=& 0. 
\end{eqnarray}

Thus, it implies that
\begin{eqnarray}
\delta W_{np3/2}^{4-3} &=& W_{4,np3/2}^{(1)} - W_{3,np3/2}^{(1)} + W_{4,np3/2}^{(2)} - W_{3,np3/2}^{(2)} \nonumber \\
   &=& 4 A^{np3/2} -(2/3) B^{np3/2} + (208/21) C^{np3/2} \nonumber \\ && + (11/3) g_I^2 \frac{|\langle J_{np1/2} || T^{(1)}_e || J_{np3/2} \rangle|^2} {E_{np3/2}-E_{np1/2}} \nonumber \\
  && - (33/2)\sqrt{1/15} g_I Q \langle J_{np1/2} || T^{(1)}_e || J_{np3/2} \rangle \nonumber \\ && \times 
  \frac { \langle J_{np1/2} || T^{(2)}_e || J_{np3/2} \rangle} {E_{np3/2}-E_{np1/2}}, \nonumber \\ 
\delta W_{np3/2}^{5-4} &=& W_{5,np3/2}^{(1)} - W_{4,np3/2}^{(1)} + W_{5,np3/2}^{(2)} - W_{4,np3/2}^{(2)} \nonumber \\
   &=& 5 A^{np3/2} -(5/24) B^{np3/2} - (65/6) C^{np3/2} \nonumber \\ && + (5/6) g_I^2  \frac{|\langle J_{np1/2} || T^{(1)}_e || J_{np3/2} \rangle|^2} {E_{np3/2}-E_{np1/2}} \nonumber \\
   && +10 \sqrt{3/5}  g_I Q \langle J_{np1/2} || T^{(1)}_e || J_{np3/2} \rangle \nonumber \\ && \times
 \frac{ \langle J_{np1/2} || T^{(2)}_e || J_{np3/2} \rangle} {E_{np3/2}-E_{np1/2}}, \nonumber \\
 \delta W_{np3/2}^{6-5} &=& W_{6,np3/2}^{(1)} - W_{5,np3/2}^{(1)} + W_{6,np3/2}^{(2)} - W_{5,np3/2}^{(2)} \nonumber \\
   &=& 6 A^{np3/2} + (2/3) B^{np3/2} + (16/3) C^{np3/2} \nonumber \\ && - (9/2) g_I^2 \frac{|\langle J_{np1/2} || T^{(1)}_e || J_{np3/2} \rangle|^2} {E_{np3/2}-E_{np1/2}} \nonumber \\
  && - (9/2) \sqrt{3/5} g_I Q \langle J_{np1/2} || T^{(1)}_e || J_{np3/2} \rangle \nonumber \\ && \times 
  \frac { \langle J_{np1/2} || T^{(2)}_e || J_{np3/2} \rangle} {E_{np3/2}-E_{np1/2}}, \nonumber \\
  \text{and} \ \ \ \  && \nonumber \\
   \delta W_{np3/2}^{6-3} &=& W_{6,np3/2}^{(1)} - W_{3,np3/2}^{(1)}  \nonumber \\
   &=& 15 A^{np3/2} - (5/24) B^{np3/2}  \nonumber \\ && + (185/42) C^{np3/2}.
 \end{eqnarray}

\section{$[np^6] \ nd ^2D_{3/2}$ state}
Since $J=3/2$ and $I=9/2$, it follows $F=3,4,5,6$ for which we have 
\begin{eqnarray}
W_{3,nd3/2}^{(1)} &=& -(33/4) A^{nd3/2} + (11/24) B^{nd3/2} \nonumber \\ && - (143/42) C^{nd3/2}, \nonumber \\ 
W_{4,nd3/2}^{(1)} &=& -(17/4) A^{nd3/2} - (5/24) B^{nd3/2} \nonumber \\ && + (13/2) C^{nd3/2}, \nonumber \\ 
W_{5,nd3/2}^{(1)} &=& (3/4) A^{nd3/2} - (5/12) B^{nd3/2} \nonumber \\ && - (13/3) C^{nd3/2}, \nonumber \\
W_{6,nd3/2}^{(1)} &=& (27/4) A^{nd3/2} + (1/4) B^{nd3/2} \nonumber \\ && + C^{nd3/2}
\end{eqnarray}
and 
\begin{eqnarray}
W_{3,nd{3/2}}^{(2)} &=& (11/10) g_I^2 \frac{|\langle J_{nd5/2} || T^{(1)}_e || J_{nd3/2} \rangle|^2} {E_{nd3/2}-E_{nd5/2}} \nonumber \\
  && - (99/20) \sqrt{1/7} g_I Q \langle J_{nd5/2} || T^{(1)}_e || J_{nd3/2} \rangle \nonumber \\ && \times 
  \frac { \langle J_{nd5/2} || T^{(2)}_e || J_{nd3/2} \rangle} {E_{nd3/2}-E_{nd5/2}}, \nonumber \\  
  W_{4,nd{3/2}}^{(2)} &=& (21/10) g_I^2 \frac{|\langle J_{nd5/2} || T^{(1)}_e || J_{nd3/2} \rangle|^2} {E_{nd3/2}-E_{nd5/2}} \nonumber \\
  && - (3/4)\sqrt{7} g_I Q \langle J_{nd5/2} || T^{(1)}_e || J_{nd3/2} \rangle \nonumber \\ && \times 
  \frac { \langle J_{nd5/2} || T^{(2)}_e || J_{nd3/2} \rangle} {E_{nd3/2}-E_{nd5/2}}, \nonumber \\  
  W_{5,nd{3/2}}^{(2)} &=& (13/5) g_I^2 \frac{|\langle J_{nd5/2} || T^{(1)}_e || J_{nd3/2} \rangle|^2} {E_{nd3/2}-E_{nd5/2}}, \nonumber \\
  W_{6,nd{3/2}}^{(2)} &=& (21/10) g_I^2 \frac{|\langle J_{nd5/2} || T^{(1)}_e || J_{nd3/2} \rangle|^2} {E_{nd3/2}-E_{nd5/2}} \nonumber \\
  && + (9/10) \sqrt{7} g_I Q \langle J_{nd5/2} || T^{(1)}_e || J_{nd3/2} \rangle \nonumber \\ && \times 
  \frac { \langle J_{nd5/2} || T^{(2)}_e || J_{nd3/2} \rangle} {E_{nd3/2}-E_{nd5/2}}. 
\end{eqnarray}

This gives us
\begin{eqnarray}
\delta W_{nd3/2}^{4-3} &=& W_{4,nd3/2}^{(1)} - W_{3,nd3/2}^{(1)} + W_{4,nd3/2}^{(2)} - W_{3,nd3/2}^{(2)} \nonumber \\
   &=& 4 A^{nd3/2} -(2/3) B^{nd3/2} + (208/21) C^{nd3/2} \nonumber \\ && + g_I^2 
   \frac{|\langle J_{nd5/2} || T^{(1)}_e || J_{nd3/2} \rangle|^2} {E_{nd3/2}-E_{nd5/2}} \nonumber \\
  && - (3/10)\sqrt{1/7} g_I Q \langle J_{nd5/2} || T^{(1)}_e || J_{nd3/2} \rangle \nonumber \\ && \times 
  \frac { \langle J_{nd5/2} || T^{(2)}_e || J_{nd3/2} \rangle} {E_{nd3/2}-E_{nd5/2}}, \nonumber \\ 
\delta W_{nd3/2}^{5-4} &=& W_{5,nd3/2}^{(1)} - W_{4,nd3/2}^{(1)} + W_{5,nd3/2}^{(2)} - W_{4,nd3/2}^{(2)} \nonumber \\
   &=& 5 A^{nd3/2} -(5/24) B^{nd3/2} - (65/6) C^{nd3/2} \nonumber \\ && + (1/2) g_I^2  
   \frac{|\langle J_{nd5/2} || T^{(1)}_e || J_{nd3/2} \rangle|^2} {E_{nd3/2}-E_{nd5/2}} \nonumber \\
   && + (3/4) \sqrt{7}  g_I Q \langle J_{nd5/2} || T^{(1)}_e || J_{nd3/2} \rangle \nonumber \\ && \times
 \frac{ \langle J_{nd5/2} || T^{(2)}_e || J_{nd3/2} \rangle} {E_{nd3/2}-E_{nd5/2}}, \nonumber \\
 \text{and} \ \ \ \ && \nonumber \\
 \delta W_{nd3/2}^{6-5} &=& W_{6,nd3/2}^{(1)} - W_{5,nd3/2}^{(1)} + W_{6,nd3/2}^{(2)} - W_{5,nd3/2}^{(2)} \nonumber \\
   &=& 6 A^{nd3/2} + (2/3) B^{nd3/2} + (16/3) C^{nd3/2} \nonumber \\ && - (1/2) g_I^2 
   \frac{|\langle J_{nd5/2} || T^{(1)}_e || J_{nd3/2} \rangle|^2} {E_{nd3/2}-E_{nd5/2}} \nonumber \\
  && + (9/10) \sqrt{7} g_I Q \langle J_{nd5/2} || T^{(1)}_e || J_{nd3/2} \rangle \nonumber \\ && \times 
  \frac { \langle J_{nd5/2} || T^{(2)}_e || J_{nd3/2} \rangle} {E_{nd3/2}-E_{nd5/2}}.
 \end{eqnarray}

\section{$[np^6] \ nd ^2D_{5/2}$ state}
For $J=5/2$ and $I=9/2$, we now have $F=2,3,4,5,6,7$. This gives many hyperfine levels for which it yields
\begin{eqnarray}
W_{2,nd5/2}^{(1)} &=& -(55/4) A^{nd5/2} + (11/24) B^{nd5/2} \nonumber \\ && - (143/42) C^{nd5/2}, \nonumber \\ 
W_{3,nd5/2}^{(1)} &=& -(43/4) A^{nd5/2} + (19/120) B^{nd5/2} \nonumber \\ && + (221/210) C^{nd5/2}, \nonumber \\ 
W_{4,nd5/2}^{(1)} &=& -(27/4) A^{nd5/2} - (1/8) B^{nd5/2} \nonumber \\ && + (559/210) C^{nd5/2}, \nonumber \\ 
W_{5,nd5/2}^{(1)} &=& -(7/4) A^{nd5/2} - (7/24) B^{nd5/2} \nonumber \\ && + (119/210) C^{nd5/2}, \nonumber \\
W_{6,nd5/2}^{(1)} &=& (17/4) A^{nd5/2} - (13/60) B^{nd5/2} \nonumber \\ && - (41/15) C^{nd5/2}, \nonumber \\
W_{7,nd5/2}^{(1)} &=& (45/4) A^{nd5/2} + (1/4) B^{nd5/2} \nonumber \\ &&  + C^{nd5/2}
\end{eqnarray}
and 
\begin{eqnarray}
W_{2,nd{5/2}}^{(2)} &=& 0, \nonumber \\
W_{3,nd{5/2}}^{(2)} &=& (11/10) g_I^2 \frac{|\langle J_{nd3/2} || T^{(1)}_e || J_{nd5/2} \rangle|^2} {E_{nd5/2}-E_{nd3/2}} \nonumber \\
  && - (99/20) \sqrt{1/7}  g_I Q \langle J_{nd3/2} || T^{(1)}_e || J_{nd5/2} \rangle \nonumber \\ && \times 
  \frac { \langle J_{nd3/2} || T^{(2)}_e || J_{nd5/2} \rangle} {E_{nd5/2}-E_{nd3/2}}, \nonumber \\
W_{4,nd{5/2}}^{(2)} &=& (21/10) g_I^2 \frac{|\langle J_{nd3/2} || T^{(1)}_e || J_{nd5/2} \rangle|^2} {E_{nd5/2}-E_{nd3/2}} \nonumber \\
  && - (3/4) \sqrt{7} g_I Q \langle J_{nd3/2} || T^{(1)}_e || J_{nd5/2} \rangle \nonumber \\ && \times 
  \frac { \langle J_{nd3/2} || T^{(2)}_e || J_{nd5/2} \rangle} {E_{nd5/2}-E_{nd3/2}}, \nonumber \\
W_{5,nd{5/2}}^{(2)} &=& (13/5) g_I^2 \frac{|\langle J_{nd3/2} || T^{(1)}_e || J_{nd5/2} \rangle|^2} {E_{nd5/2}-E_{nd3/2}}, \nonumber \\
W_{6,nd{5/2}}^{(2)} &=& (21/10) g_I^2 \frac{|\langle J_{nd3/2} || T^{(1)}_e || J_{nd5/2} \rangle|^2} {E_{nd5/2}-E_{nd3/2}} \nonumber \\
  && + (9/10) \sqrt{7} g_I Q \langle J_{nd3/2} || T^{(1)}_e || J_{nd5/2} \rangle \nonumber \\ && \times 
  \frac { \langle J_{nd3/2} || T^{(2)}_e || J_{nd5/2} \rangle} {E_{nd5/2}-E_{nd3/2}}, \nonumber \\
W_{7,nd{5/2}}^{(2)} &=& 0.  
\end{eqnarray}

This corresponds to
\begin{eqnarray}
\delta W_{nd5/2}^{3-2} &=& W_{3,nd5/2}^{(1)} - W_{2,nd5/2}^{(1)} + W_{3,nd5/2}^{(2)} - W_{2,nd5/2}^{(2)} \nonumber \\
   &=& 3 A^{nd5/2} -(3/10) B^{nd5/2} + (156/35) C^{nd5/2} \nonumber \\ && + (11/10) g_I^2 
   \frac{|\langle J_{nd3/2} || T^{(1)}_e || J_{nd5/2} \rangle|^2} {E_{nd5/2}-E_{nd3/2}} \nonumber \\
  && - (99/20)\sqrt{1/7} g_I Q \langle J_{nd3/2} || T^{(1)}_e || J_{nd5/2} \rangle \nonumber \\ && \times 
  \frac { \langle J_{nd3/2} || T^{(2)}_e || J_{nd5/2} \rangle} {E_{nd5/2}-E_{nd3/2}}, \nonumber \\ 
\delta W_{nd5/2}^{4-3} &=& W_{4,nd5/2}^{(1)} - W_{3,nd5/2}^{(1)} + W_{4,nd5/2}^{(2)} - W_{3,nd5/2}^{(2)} \nonumber \\
   &=& 4 A^{nd5/2} -(17/60) B^{nd5/2} + (169/105) C^{nd5/2} \nonumber \\ && + g_I^2 
   \frac{|\langle J_{nd3/2} || T^{(1)}_e || J_{nd5/2} \rangle|^2} {E_{nd5/2}-E_{nd3/2}} \nonumber \\
  && - (3/10)\sqrt{1/7} g_I Q \langle J_{nd3/2} || T^{(1)}_e || J_{nd5/2} \rangle \nonumber \\ && \times 
  \frac { \langle J_{nd3/2} || T^{(2)}_e || J_{nd5/2} \rangle} {E_{nd5/2}-E_{nd3/2}}, \nonumber \\ 
\delta W_{nd5/2}^{5-4} &=& W_{5,nd5/2}^{(1)} - W_{4,nd5/2}^{(1)} + W_{5,nd5/2}^{(2)} - W_{4,nd5/2}^{(2)} \nonumber \\
   &=& 5 A^{nd5/2} -(1/6) B^{nd5/2} - (44/21) C^{nd5/2} \nonumber \\ && + (1/2) g_I^2  
   \frac{|\langle J_{nd3/2} || T^{(1)}_e || J_{nd5/2} \rangle|^2} {E_{nd5/2}-E_{nd3/2}} \nonumber \\
   && + (3/4) \sqrt{7}  g_I Q \langle J_{nd3/2} || T^{(1)}_e || J_{nd5/2} \rangle \nonumber \\ && \times
 \frac{ \langle J_{nd3/2} || T^{(2)}_e || J_{nd5/2} \rangle} {E_{nd5/2}-E_{nd3/2}}, \nonumber \\
 \delta W_{nd5/2}^{6-5} &=& W_{6,nd5/2}^{(1)} - W_{5,nd5/2}^{(1)} + W_{6,nd5/2}^{(2)} - W_{5,nd5/2}^{(2)} \nonumber \\
   &=& 6 A^{nd5/2} + (3/40) B^{nd5/2} - (33/10) C^{nd5/2} \nonumber \\ && - (1/2) g_I^2 
   \frac{|\langle J_{nd3/2} || T^{(1)}_e || J_{nd5/2} \rangle|^2} {E_{nd5/2}-E_{nd3/2}} \nonumber \\
  && + (9/10) \sqrt{7} g_I Q \langle J_{nd3/2} || T^{(1)}_e || J_{nd5/2} \rangle \nonumber \\ && \times 
  \frac { \langle J_{nd3/2} || T^{(2)}_e || J_{nd5/2} \rangle} {E_{nd5/2}-E_{nd3/2}}, \nonumber \\
 \delta W_{nd5/2}^{7-6} &=& W_{7,nd5/2}^{(1)} - W_{6,nd5/2}^{(1)} + W_{7,nd5/2}^{(2)} - W_{6,nd5/2}^{(2)} \nonumber \\
   &=& 7 A^{nd5/2} + (7/15) B^{nd5/2} + (56/15) C^{nd5/2} \nonumber \\ && - (21/10) g_I^2 
   \frac{|\langle J_{nd3/2} || T^{(1)}_e || J_{nd5/2} \rangle|^2} {E_{nd5/2}-E_{nd3/2}} \nonumber \\
  && - (9/10) \sqrt{7} g_I Q \langle J_{nd3/2} || T^{(1)}_e || J_{nd5/2} \rangle \nonumber \\ && \times 
  \frac { \langle J_{nd3/2} || T^{(2)}_e || J_{nd5/2} \rangle} {E_{nd5/2}-E_{nd3/2}}, \nonumber \\ 
  \text{and} \ \ \ \ && \nonumber \\
  \delta W_{nd5/2}^{7-2} &=& W_{7,nd5/2}^{(1)} - W_{2,nd5/2}^{(1)} \nonumber \\
   &=& 25 A^{nd5/2} - (5/24) B^{nd5/2} \nonumber \\ && + (185/42) C^{nd5/2}.
  \end{eqnarray}

\end{document}